\documentclass[aps,showpacs,preprint,preprintnumber,nofootinbib,amsmath,amssymb,ascmac,12pt]{revtex4}
\usepackage{bm}
\usepackage{color}
\usepackage[dvipdfm]{graphicx}
\newcommand{\eq}[1]{(\ref{#1})}
\begin{document}
\title{
What wormhole is traversable?\\
{\it
--
A case of a wormhole supported by a spherical thin shell
--
}
}
\author{
${}^{1,2}$Ken-ichi Nakao\footnote{E-mail:knakao@sci.osaka-cu.ac.jp},
${}^{2}$Tatsuya Uno\footnote{E-mail:uno@sci.osaka-cu.ac.jp}
and
${}^{3}$Shunichiro Kinoshita\footnote{E-mail:kinosita@sci.osaka-cu.ac.jp}
}
\affiliation{
${}^{1}$DAMTP, Centre for Mathematical Sciences, University of Cambridge, Wilberforce
Road, Cambridge CB3 0WA, United Kingdom
\\
${}^{2}$Department of Mathematics and Physics,
Graduate School of Science, Osaka City University,
3-3-138 Sugimoto, Sumiyoshi, Osaka 558-8585, Japan
\\
${}^{3}$Osaka City University Advanced Mathematical Institute, 
Osaka City University,
3-3-138 Sugimoto, Sumiyoshi, Osaka 558-8585, Japan
\vspace{1cm}
}
\begin{abstract}
We analytically explore the effect of falling matter on a spherically symmetric wormhole 
supported by a spherical shell composed of exotic matter located at its throat. 
The falling matter is assumed to be also a thin spherical shell concentric with the 
shell supporting the wormhole, 
and its self-gravity is completely taken into account. 
We treat these spherical thin shells by Israel's formalism 
of metric junction. When the falling spherical shell goes through the wormhole, 
it necessarily collides with the  shell supporting the wormhole. 
To treat this collision, we assume the interaction between these shells 
is only gravity. We show the conditions on the parameters that characterize 
this model in which the wormhole persists after the spherical shell goes 
through it.

\end{abstract}

\preprint{OCU-PHYS 386}
\preprint{AP-GR 106}

\date{\today}
\maketitle

\section{Introduction}

The wormhole is a tunnel-like spacetime structure by which a shortcut or travel to disconnected 
world is possible.  
Active theoretical studies of this fantastic subject began by an influential paper written by 
Morris, Thorne and Yurtsever\cite{MTY1988} and Morris and Thorne\cite{MT1988}. 
The earlier works are shown in the book written by 
Visser\cite{Visser1995} and review paper by Lobo\cite{Lobo2008}.  
However, it is not trivial what is the mathematically rigorous and physically reasonable 
definition of wormhole in general situation, although we may find a wormhole structure in 
each individual case. Hayward gave an elegant definition of wormhole by using trapping 
horizon and showed that the violation of the 
null energy condition is a necessary condition for the existence of the wormhole 
in the framework of general relativity, where the null energy condition is  
$T_{\mu\nu}k^\mu k^\nu\geq0$ for any null 
vector $k^\mu$\cite{Hayward1999,Hayward2009}.  

The exotic matter is necessary to make a wormhole, but where is an exotic matter? 
In Refs.\cite{MTY1988} and \cite{MT1988}, the authors discussed possibilities of quantum effects. 
Alternatively, 
such an exotic matter is often discussed in the context of cosmology. 
The phantom energy, 
whose equation of state is $p=w\rho$ with $w<-1$ and positive energy
density $\rho>0$, 
does not satisfy the null energy condition, and a few researches showed the 
possibility of the wormhole supported phantom-like matter\cite{Sushkov2005,Lobo2005-1,Lobo2005-2}.  
Recently, theoretical studies from observational point of view on a compact object 
made of the exotic matter,  possibly wormholes, have also 
reported\cite{Nakajima-Asada2012,Tsukamoto-Harada2012}.

It is very important to study the stability of wormhole model in order to know whether it is traversable. 
The stability against linear perturbations is a necessary condition for the traversable wormhole, 
but it is insufficient. 
The investigation of non-linear dynamical situation is necessary, and there are a few studies 
in this direction\cite{Shinkai-Hayward,Hayward-Koyama2004,Koyama-Hayward2004}. 
In this paper, we study the condition that a wormhole persists even if it experiences non-linear  
disturbances. In our model, the wormhole is assumed to be supported by a spherical 
thin shell composed of the exotic matter, and hence the wormhole itself is also assumed to be 
spherically symmetric. The largest merit of a spherical thin shell 
wormhole is the finite number of its dynamical degrees of 
freedom. By virtue of this merit, we can analyze this model analytically even in 
highly dynamical cases. 
The thin shell wormhole was first devised by Visser\cite{Visser1989}, and then its stability 
against linear perturbations was investigated by Poisson and Visser\cite{Poisson-Visser1995}. 
Recently, the linear stability of the thin shell wormhole in more general situation has been 
investigated by Garcia, Lobo and Visser\cite{GLV2012}. 

In this paper, we consider a situation in which a spherical thin shell 
concentric with a wormhole supported by 
another thin shell enters the wormhole. These spherical shells are treated 
by Israel's formulation of metric 
junction\cite{Israel}. When the shell goes through the wormhole, it necessarily collides 
with the shell supporting the wormhole. The collision between thin shells has already studied by 
several researchers\cite{NIS1999,IN1999,LMW2002}, and we follow them. 
Then, we show the condition that the wormhole persists after a spherical shell 
passes the wormhole. 

This paper is organized as follows. In Sec.~II, we derive the equations of motion for 
the spherical shell supporting the wormhole and the other spherical shell falling 
into the wormhole, in accordance with Israel's formalism of metric junction. 
In Sec.~III, we derive a static solution of thin shell wormhole which is 
the initial condition. In Sec.~IV, we reveal the condition 
that a shell falls from infinity and goes through the wormhole. 
In Sec.~V, we study the motion of the shells and the change in the gravitational mass 
of the wormhole after collision. In Sec. VI, we show the condition 
that the wormhole persists after the shell goes through it. 
Sec. VII is devoted to summary and discussion. 

In this paper, we adopt the geometrized unit in which the speed of light and 
Newton's gravitational constant are one. 

\section{Equation of motions for spherical shells}

\begin{figure}[b]
\begin{center}
\includegraphics[width=0.3\textwidth]{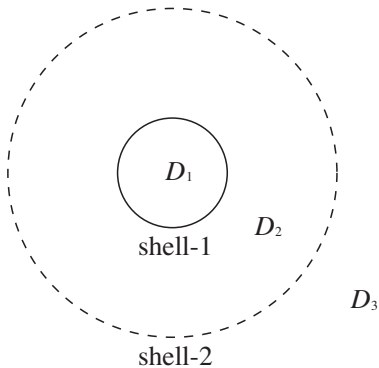}
\caption{\label{initial}
The initial configuration is depicted. 
}
\end{center}
\end{figure}

We consider two concentric spherical shells which are infinitesimally thin. 
The trajectories of these shells in the spacetime are timelike hypersurfaces: The 
inner hypersurface is denoted by $\Sigma_1$, and the outer hypersurface 
is denoted by $\Sigma_2$. These hypersurfaces divide a domain of the spacetime into 
three domains: The innermost domain is denoted by $D_1$, the middle one is 
denoted by $D_2$, and the outermost one is denoted by $D_3$. 
We also call $\Sigma_1$ and $\Sigma_2$ the shell-1 and the shell-2, respectively. 
This configuration is depicted in Fig.~\ref{initial}. 

By the symmetry of this system, the geometry of the domain $D_i$ ($i=1,2,3$) 
is described by the Schwarzschild solution whose line element is given by
\begin{eqnarray}
ds^2=-f_i (r)dt_i^2+\frac{1}{f_i(r)}dr^2
+r^2\left(d\theta^2+\sin^2\theta d\phi^2\right)  \label{RN}
\end{eqnarray}
with
\begin{equation}
f_i(r)=1-\frac{2M_i}{r},
\end{equation}
where $M_i$ is the mass parameter. We should note that the coordinate $t_i$ is 
not continuous across the shells, whereas $r$, $\theta$ and $\phi$ are continuous across 
the shells. 

The location of the horizon is given by a solution of the equation $f_i(r)=0$ as 
\begin{equation}
r=r_{i}\equiv 2M_i.
\end{equation} 
The positive root exists if and only if $M_i$ is positive. 

Since finite energy and finite momentum concentrate on the infinitesimally 
thin domains, the stress-energy tensor diverges on these shells. This means 
that these shells are categorized into the so-called curvature polynomial singularity 
through the Einstein equations. Even though $\Sigma_A$ ($A=1,2$) are spacetime 
singularities, we can derive the equation of 
motion for each spherical shell which is consistent with the Einstein equations 
by so-called Israel's formalism. 

Let us cover the neighborhood of one singular hypersurface $\Sigma_A$
by a Gaussian normal coordinate $\lambda$, 
where $\partial/\partial\lambda$ is 
a unit vector normal to $\Sigma_A$ and directs from 
$D_A$ to $D_{A+1}$. Then, the sufficient condition 
to apply Israel's formalism is that the stress-energy tensor is written in the form
\begin{equation}
T_{\mu\nu}=S_{\mu\nu}\delta(\lambda-\lambda_A)
\end{equation}
where $\Sigma_A$ is located at $\lambda=\lambda_A$, 
$\delta(x)$ is Dirac's delta function, 
and $S_{\mu\nu}$ is the surface stress-energy tensor on  $\Sigma_A$. 

The junction condition of the metric tensor is given as follows.  
We impose that the metric tensor $g_{\mu\nu}$ is continuous across $\Sigma_A$.  
Hereafter, we denote the unit normal vector of $\Sigma_A$ 
by $n^\mu$ instead of $\partial/\partial\lambda$. 
The intrinsic metric of $\Sigma_A$ is given by
\begin{equation}
h_{\mu\nu}=g_{\mu\nu}-n_\mu n_\nu,
\end{equation}
and the extrinsic curvature is defined by
\begin{equation}
K^{(i)}_{\mu\nu}=-h^{\alpha}{}_\mu h^{\beta}{}_\nu \nabla^{(i)}_\alpha n_\beta,
\end{equation}
where $\nabla^{(i)}_\alpha$ is the covariant derivative with respect to the metric in the 
domain $D_i$. This extrinsic curvature describes how $\Sigma_A$ is embedded into 
the domain $D_i$. In accordance with Israel's formalism, the Einstein equations lead to
\begin{equation}
K^{(A+1)}_{\mu\nu}-K^{(A)}_{\mu\nu}=8\pi\left(S_{\mu\nu}
-\frac{1}{2}h_{\mu\nu}{\rm tr}S\right),
\label{j-con-0}
\end{equation}
where ${\rm tr}S$ is the trace of $S_{\mu\nu}$.  Equation (\ref{j-con-0}) gives the condition 
of the metric junction. 

By the spherical symmetry, the surface stress-energy tensors of the shells 
should be the perfect fluid type; 
\begin{equation}
S_{\mu\nu}=\sigma u_\mu u_\nu+P(h_{\mu\nu}+u_\mu u_\nu),
\end{equation}
where $\sigma$ and $P$ are the energy per unit area and the pressure 
on $\Sigma_A$, respectively, 
and $u^\mu$ is the 4-velocity. 

By the spherical symmetry, 
the motion of the shell-$A$ is described in the form of $t_i=T_{A,i}(\tau)$ and 
$r=R_A(\tau)$, where $i=A$ or $i=A+1$,
that is to say, $i$ represents one of two domains divided by the shell-$A$, 
and $\tau$ is the proper time of the shell.  
The 4-velocity is given by
\begin{equation}
u^\mu=\left(\dot{T}_{A,i},\dot{R}_A,0,0\right),
\end{equation}
where a dot means the derivative with respect to $\tau$. 
Then, $n_\mu$ is given by
\begin{equation}
n_\mu=\left(-\dot{R}_A,\dot{T}_{A,i},0,0\right).
\end{equation}
Together with $u^\mu$ and $n^\mu$, the following unit vectors form an orthonormal frame;
\begin{eqnarray}
e_{(\theta)}^\mu&=&\left(0,0,\frac{1}{r},0\right), \\
e_{(\phi)}^\mu&=&\left(0,0,0,\frac{1}{r\sin\theta}\right).
\end{eqnarray}

The extrinsic curvature is obtained as
\begin{eqnarray}
K_{\mu\nu}^{(i)}u^\mu u^\nu&=&\frac{1}{f_{i}\dot{T}_{A,i}}\left(\ddot{R}_A+\frac{f'_i(R_A)}{2}\right), \\
K^{(i)}_{\mu\nu}e_{(\theta)}^\mu e_{(\theta)}^\nu&=&
K_{\mu\nu}^{(i)}e_{(\phi)}^\mu e_{(\phi)}^\nu=-n^a\partial_a \ln r|_{D_i}=-\frac{f_i(R_A)}{R_A}\dot{T}_{A,i}
\label{th-th-comp}
\end{eqnarray}
and the other components vanish, 
where a prime means a derivative with respect to its argument. 
By the normalization condition $u^\mu u_\mu=-1$, we have
\begin{equation}
\dot{T}_{A,i}= \pm\frac{1}{f_i(R_A)}\sqrt{\dot{R}_A^2+f_i(R_A)}~. \label{t-dot}
\end{equation}
Substituting the above equation into Eq.~(\ref{th-th-comp}), we have
\begin{equation}
K_{\mu\nu}^{(i)}e_{(\theta)}^\mu e_{(\theta)}^\nu=\mp \frac{1}{R_A}\sqrt{\dot{R}_A^2+f_i(R_A)}.
\end{equation}

From the $u$-$u$ component of Eq.~\eq{j-con-0}, we obtain the following relations.
\begin{equation}
\frac{d(\sigma_A R_A^2)}{d\tau}+P_A\frac{dR_A^2}{d\tau}=0. \label{E-con}
\end{equation}
Here, we assume the following equation of state
\begin{equation}
P_A=w_A \sigma_A, \label{EOS}
\end{equation}
where $w_A$ is constant. Substituting Eq.~(\ref{EOS}) into Eq.~(\ref{E-con}), we obtain
\begin{equation}
\sigma_A \propto R_A^{-2(w_A+1)}. \label{sigma}
\end{equation}

\subsection{The shell-1: Initially inner shell}

\begin{figure}
\begin{center}
\includegraphics[width=0.5\textwidth]{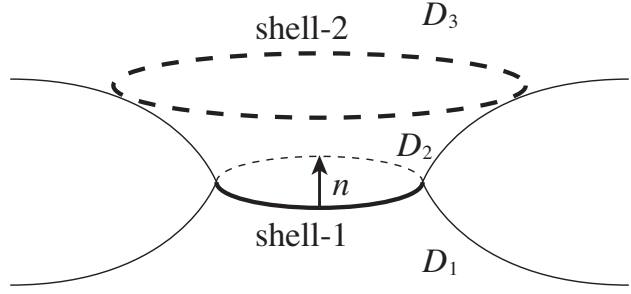}
\caption{\label{fig:wh}
The shell-1 forms the wormhole structure. 
}
\end{center}
\end{figure}

We assume that the domains $D_1$ and $D_2$ form a 
wormhole structure by the shell-1. This means that $n^a\partial_a\ln r |_{D_1}<0$ and 
$n^a\partial_a\ln r |_{D_2}>0$ (see Fig.~\ref{fig:wh}), and we have 
\begin{equation}
K_{\mu\nu}^{(1)}e_{(\theta)}^\mu e_{(\theta)}^\nu= +\frac{1}{R_1}\sqrt{\dot{R}_1^2+f_1}
~~~~~{\rm and}~~~~~
K_{\mu\nu}^{(2)}e_{(\theta)}^\mu e_{(\theta)}^\nu= -\frac{1}{R_1}\sqrt{\dot{R}_1^2+f_2}.
\label{K-1}
\end{equation}
Here, note that Eq.~(\ref{K-1}) implies 
$\dot{T}_{1,1}$ is negative, whereas $\dot{T}_{1,2}$ is positive. 
Hence, the direction of the time coordinate basis vector in $D_1$ is opposite with that in $D_2$.

From $\theta$-$\theta$ component of Eq.~\eq{j-con-0}, we obtain the following relations. 
\begin{equation}
\sqrt{\dot{R}_1^2+f_2(R_1)}+\sqrt{\dot{R}_1^2+f_1(R_1)}= -4\pi \sigma_1 R_1. \label{j-con-1}
\end{equation}
Equation (\ref{j-con-1}) is satisfied only if $\sigma_1$ is negative, 
and hence we assume so. Here, we introduce a new positive variable defined by
\begin{equation}
m_1\equiv-4\pi\sigma_1 R_1^2.
\end{equation}
 From Eq.~(\ref{sigma}), we have
\begin{equation}
m_1=\mu R_1^{-2w_1},
\end{equation}
where $\mu$ is a positive constant, and, for notational simplicity, hereafter we denote $w_1$ by $w$. 

Let us rewrite Eq.~\eq{j-con-1} into the form of the energy equation for the shell-1. 
First, we write it in the form
\begin{equation}
\sqrt{\dot{R}_1^2+f_2(R_1)}=-\sqrt{\dot{R}_1^2+f_1(R_1)}+\frac{m_1}{R_1},
\label{j-con-1-1}
\end{equation}
and then take a square of the both sides of the above equation:
\begin{equation}
\dot{R}_1^2+f_{2}(R_1)=\dot{R}_1^2+f_1(R_1)+\left(\frac{m_1}{R_1}\right)^2
-\frac{2m_1}{R_1}\sqrt{\dot{R}_1^2+f_1(R_1)}.
\label{j-con-1-2}
\end{equation}
Furthermore, we rewrite the above equation in the form
\begin{equation}
\sqrt{\dot{R}_1^2+f_1(R_1)}=\frac{R_1}{2m_1}\left[f_1(R_1)-f_2(R_1)
+\left(\frac{m_1}{R_1}\right)^2\right].
\label{j-con-1-3}
\end{equation}
By taking a square of the both sides of the above equation, we have
\begin{equation}
\dot{R}_1^2+V_1(R_1)=0, 
\label{energy-eq}
\end{equation}
where
\begin{eqnarray}
V_1(r)&\equiv&f_1(r)-\left(\frac{r}{2m_1}\right)^2 
\left[
f_1(r)-f_2(r)+\left(\frac{m_1}{r}\right)^2
\right]^2 \cr
&&\cr
&=&1-\left(\frac{M_2-M_1}{m_1}\right)^2-\frac{M_1+M_2}{r}-\left(\frac{m_1}{2r}\right)^2 \cr
&& \cr
&=&1-{\cal E}r^{4w}-\frac{2M_{\rm wh}}{r}-\left(\frac{\mu}{2}\right)^2 r^{-2(2w+1)},
\label{potential}
\end{eqnarray}
where 
\begin{equation}
{\cal E}\equiv \left(\frac{M_2-M_1}{\mu}\right)^2
~~~~~~{\rm and}~~~~~~M_{\rm wh} \equiv \frac{M_1+M_2}{2}.
\end{equation}
Equation (\ref{energy-eq}) is regarded as the energy equation for the shell-1. 
The function $V_1$ corresponds to the effective potential. In the allowed domain 
for the motion of the shell-1, an inequality $V_1\leq0$ should hold. 
But, this inequality is not a sufficient condition of the allowed region. 

The left hand side of Eq.~(\ref{j-con-1-1}) is non-negative, and hence the right hand side of it 
should also be non-negative. Then, substituting Eq.~(\ref{j-con-1-3}) into the right hand side of 
Eq.~(\ref{j-con-1-1}), we have
\begin{eqnarray}
0&\leq& -\sqrt{\dot{R}_1^2+f_1(R_1)}+\frac{m_1}{R_1}
=-\frac{R_1}{2m_1}\left[f_1(R_1)-f_2(R_1)+\left(\frac{m_1}{R_1}\right)^2\right]+\frac{m_1}{R_1} \cr
&=&\frac{m_1}{2R_1}-\frac{M_2-M_1}{m_1}.
\end{eqnarray}
Further manipulation leads to
\begin{equation}
R_1^{-(4w+1)}\geq \frac{2}{\mu^2}(M_2-M_1).
\end{equation}
By the similar argument, we obtain
\begin{equation}
-\sqrt{\dot{R}_1^2+f_2(R_1)}+\frac{m_1}{R_1}\geq0.
\end{equation}
Then, by the similar procedure, we have
\begin{equation}
R_1^{-(4w+1)}\geq \frac{2}{\mu^2}(M_1-M_2).
\end{equation}
Hence, we have
\begin{equation}
R_1^{-(4w+1)}\geq \frac{2}{\mu^2}|M_2-M_1|=\frac{2}{\mu}\sqrt{\cal E}.
\end{equation}
Finally, we obtain the following constraint; 
\begin{equation}
R_1\leq \left(\frac{\mu}{2\sqrt{\cal E}}\right)^{1\over{4w+1}}~~~~~~{\rm for}~~4w+1>0
\label{ad-con-1}
\end{equation}
or
\begin{equation}
R_1\geq \left(\frac{\mu}{2\sqrt{\cal E}}\right)^{-{1\over{4w+1}}}~~~~~{\rm for}~~4w+1<0.
\label{ad-con-2}
\end{equation}
In order to find the allowed domain for the motion of the shell-1, 
we need to take into account the constraint (\ref{ad-con-1}) or (\ref{ad-con-2}) 
in addition to the condition $V_1\leq0$. 

\subsection{The shell-2: Initially outer shell}

For simplicity, we assume that the outer shell (shell-2) is composed of dust, i.e., $w_2=0$. 
The proper mass of the shell-2 is defined by
\begin{equation}
m_2\equiv 4\pi\sigma_2 R_2^2.
\end{equation}
By Eq.~(\ref{sigma}), we find that $m_2$ is constant. 
We assume that $\sigma_2$ takes any value except for the trivial case $\sigma_2=0$, 
and hence $m_2$ can take any value except for the trivial case $m_2=0$.  

We assume the wormhole structure does not exist around the shell-2. Hence, 
the extrinsic curvature of the shell-2 is given by
\begin{equation}
K_{\mu\nu}^{(2)}e_{(\theta)}^\mu e_{(\theta)}^\nu= -\frac{1}{R_2}\sqrt{\dot{R}_2^2+f_2(R_2)}
~~~~~{\rm and}~~~~~
K_{\mu\nu}^{(3)}e_{(\theta)}^\mu e_{(\theta)}^\nu= -\frac{1}{R_2}\sqrt{\dot{R}_2^2+f_3(R_2)}.
\end{equation}
By using the above result, the $\theta$-$\theta$ component of the junction condition leads to
\begin{equation}
\sqrt{\dot{R}_2^2+f_3(R_2)}-\sqrt{\dot{R}_2^2+f_2(R_2)}= -\frac{m_2}{R_2}. \label{j-con-out-1}
\end{equation}

In the case of $m_2>0$, we find from the above equation that $f_2(R_2)> f_3(R_2)$, or equivalently, 
$M_3 > M_2$. 
From the above equation, we have
\begin{equation}
\sqrt{\dot{R}_2^2+f_3(R_2)}=\sqrt{\dot{R}_2^2+f_2(R_2)} -\frac{m_2}{R_2}. \label{j-con-out-2}
\end{equation}
Since the left hand side of the above equation is non-negative, the following inequality should be 
satisfied. 
\begin{equation}
\sqrt{\dot{R}_2^2+f_2(R_2)} -\frac{m_2}{R_2}\geq0. \label{j-con-out-3}
\end{equation}
By taking the square of the both sides of Eq.~(\ref{j-con-out-2}), we have
\begin{equation}
\sqrt{\dot{R}_2^2+f_2(R_2)}=\frac{M_3-M_2}{m_2}+\frac{m_2}{2R_2}.\label{j-con-out-4}
\end{equation}
Substituting the above result into the left hand side of Eq.~(\ref{j-con-out-3}), we have 
\begin{equation}
R_2\geq  \frac{m_2^2}{2(M_3-M_2)}. \label{r-cons-out+}
\end{equation}
In the case of $m_2<0$, we find from Eq.~(\ref{j-con-out-1}) that $f_2(R_2)<f_3(R_2)$, or equivalently, 
$M_3 < M_2$. 
From Eq.~(\ref{j-con-out-1}), we have
\begin{equation}
\sqrt{\dot{R}_2^2+f_2(R_2)}=\sqrt{\dot{R}_2^2+f_3(R_2)} +\frac{m_2}{R_2}. \label{j-con-out-2-2}
\end{equation}
Since the left hand side of the above equation is non-negative, the following inequality should be 
satisfied. 
\begin{equation}
\sqrt{\dot{R}_2^2+f_3(R_2)} +\frac{m_2}{R_2}\geq0. \label{j-con-out-3-2}
\end{equation}
By taking the square of the both sides of Eq.~(\ref{j-con-out-2-2}), we have
\begin{equation}
\sqrt{\dot{R}_2^2+f_2(R_2)}=\frac{M_3-M_2}{m_2}-\frac{m_2}{2R_2}.\label{j-con-out-4-2}
\end{equation}
Substituting the above result into the left hand side of Eq.~(\ref{j-con-out-3-2}), we have 
\begin{equation}
R_2\geq \frac{m_2^2}{2(M_2-M_3)}. \label{r-cons-out-}
\end{equation}
From Eqs.~(\ref{r-cons-out+}) and (\ref{r-cons-out-}), we have
\begin{equation}
R_2\geq R_{\rm b}\equiv \frac{m_2^2}{2|M_2-M_3|}. \label{r-cons-out}
\end{equation}

By taking a square of both sides of Eq.~(\ref{j-con-out-4}),  
we obtain an energy equation for the shell-2, 
\begin{equation}
\dot{R}_2^2+V_2(R_2)=0, \label{e-eq-2}
\end{equation} 
where
\begin{equation}
V_2(r)=1-E-\frac{2M_{\rm d}}{r}-\left(\frac{m_2}{2r}\right)^2, \label{V2-def}
\end{equation}
with
\begin{equation}
E\equiv \left(\frac{M_3-M_2}{m_2}\right)^2~~~~~~{\rm and}~~~~~~
M_{\rm d}\equiv\frac{1}{2}(M_2+M_3).
\end{equation}
Note that $E$ is a constant which corresponds to the square of the specific energy of the shell-2. 
The allowed domain for the motion of the shell-2 satisfies $V_2\leq0$ and  
Eq.~(\ref{r-cons-out}) as long as $R_{\rm b}\geq 2M_3$.\footnote{Equation~(\ref{r-cons-out}) is derived by using $u^t_2$ is positive, but $u^t_2$ can change its sign within the black hole $R_2<2M_3$. Hence, if $R_{\rm b}$ is smaller than $2M_3$, Eq.~(\ref{r-cons-out}) looses its validity, and thus the allowed domain for the motion of the shell-2 is determined by the only condition $V_2\leq0$. }

\section{Static wormhole solution}

We consider a situation in which the wormhole supported by the shell-1 is initially static. 
For simplicity, we assume the symmetric wormhole, i.e., 
${\cal E}=0$ or equivalently, $M_1=M_2=M_{\rm wh}$. In this case, the analysis becomes very simple. 
In order that the wormhole structure is static, the areal radius $R_1=a$ of the shell-1 should satisfy 
$V_1(a)=0=V'_1(a)$. Furthermore, in order that this structure is stable, 
$V_1''(a)>0$ should be satisfied. 
These conditions lead to 
\begin{equation}
a^{4w+2}-2M_{\rm wh}a^{4w+1}-\left(\frac{\mu}{2}\right)^2=0, \label{V=0}
\end{equation}
\begin{equation}
2M_{\rm wh}a^{4w+1}+\frac{\mu^2}{2}(2w+1)=0 \label{dV=0}
\end{equation}
and
\begin{equation}
4M_{\rm wh}a^{4w+1}+\frac{\mu^2}{2}(4w+3)(2w+1)<0. \label{ddV>0}
\end{equation}

From Eq.~(\ref{dV=0}), we have
\begin{equation}
a^{4w+1}=-\frac{\mu^2}{4M_{\rm wh}}(2w+1). \label{a-sol-1}
\end{equation}
Since $a$ should be positive, we find that both $M_{\rm wh}>0$ and $2w+1<0$ should be satisfied, or 
both $M_{\rm wh}<0$ and $2w+1>0$ should be satisfied. 
Substituting Eq.~(\ref{a-sol-1}) into the left hand side of Eq.~(\ref{ddV>0}), we have
\begin{equation}
\frac{\mu^2}{2}(2w+1)(4w+1)<0.
\end{equation}
The above inequality implies $-1/2<w<-1/4$. 
Hence, the static symmetric wormhole is stable, only if 
\begin{equation}
M_{\rm wh}<0~~~~{\rm and}~~~~-\frac{1}{2}<w<-\frac{1}{4}. \label{wh-con}
\end{equation}

Substituting Eq.~(\ref{a-sol-1}) into Eq.~(\ref{V=0}), we have
\begin{equation}
\mu^2=\left(-\frac{4w+1}{4}\right)^{4w+1}\left(-\frac{4M_{\rm wh}}{2w+1}\right)^{4w+2}.
\label{mu-sol}
\end{equation}
Substituting the above equation into Eq.~(\ref{a-sol-1}), we have
\begin{equation}
a^{4w+1}=\left(\frac{4w+1}{2w+1}M_{\rm wh}\right)^{4w+1},
\end{equation}
and hence
\begin{equation}
a=F(w)M_{\rm wh}, \label{a-sol}
\end{equation}
where
\begin{equation}
F(w)\equiv\frac{4w+1}{2w+1}. \label{F-def}
\end{equation}
The factor $F(w)$ is monotonically increasing in the domain $-1/2<w<-1/4$, and 
$F(w)\rightarrow -\infty$ for $w\rightarrow-1/2$, whereas $F(w)\rightarrow 0$ for $w\rightarrow-1/4$.

\section{The condition of the entrance to the wormhole}

We show the condition that the shell-2 enters the wormhole supported by the shell-1. 
The allowed domain for the motion of the shell-2 is determined by the conditions 
(\ref{r-cons-out}) and $V_2\leq0$.  

\begin{figure}
\begin{center}
\includegraphics[width=0.8\textwidth]{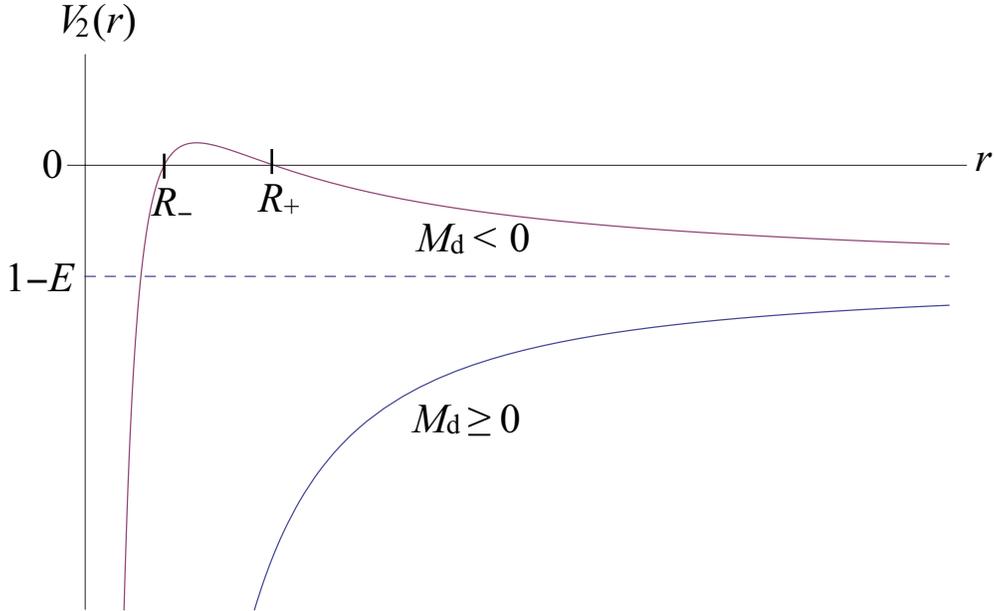}
\caption{\label{V2}
The effective potential of the shell-2. 
}
\end{center}
\end{figure}

The shell-2 is assumed to come from the spatial infinity. By this assumption, 
$E>1$ or $E=1$ with $M_{\rm d}\geq0$ should hold from the condition that $V_2(r)<0$ 
for sufficiently large $r$. If $M_{\rm d}\geq0$, we easily see that $V_2(r)<0$ always holds for 
$E\geq1$, and hence the shell-2 can enter the wormhole 
(see Fig. \ref{V2}).

In the case of $M_{\rm d}<0$ (hence $E$ should be larger than unity), 
the equation $V'_2(r)=0$ has a positive root 
\begin{equation}
r=r_{\rm m}\equiv-\frac{m_2^2}{4M_{\rm d}}, 
\end{equation}
where a prime represents a derivative with respect to the argument. 
Furthermore, if  $V_2(r_{\rm m})>0$, or equivalently, 
\begin{equation}
1+\frac{4}{m_2^2}M_2M_3\geq0, \label{Dis}
\end{equation}
the shell-2 falling from infinity may be eventually prevented from the entrance to the wormhole 
by the potential barrier. 
If the inequality in Eq.~(\ref{Dis}) holds, the equation $V_2(r)=0$ has two positive roots given by
\begin{equation}
r=R_\pm\equiv\frac{1}{E-1}\left[
-M_{\rm d}\pm\sqrt{M_{\rm d}^2-\frac{m_2^2}{4}(E-1)}
\right].
\label{rho}
\end{equation}
If the equality of Eq.~(\ref{Dis}) holds, $R_+$ agrees with $R_-$. 
The shell-2 cannot enter the domain $R_2<R_+$ as long as Eq.~(\ref{Dis}) is satisfied. 
Hence, if Eq.~(\ref{Dis}) is satisfied, $R_+<a$ should be satisfied so 
that the shell-2 enters the wormhole.  

Let us investigate $V_2(R_{\rm b})$, where $R_{\rm b}$ is defined in Eq.~(\ref{r-cons-out}). 
We have
\begin{equation}
V_2(R_{\rm b})=\left\{
\begin{array}{ll}
1-4M_3(M_3-M_2)/m_2^2
&~~~~~~ \mbox{$M_3>M_2$,} \\
1-4M_2(M_2-M_3)/m_2^2
&~~~~~~ \mbox{$M_3\leq M_2$.}
\end{array}
\right.
\end{equation}
Let us consider two cases $M_3\leq 0$ and $M_3 >0$, separately. 
In the former case, $V_2(R_{\rm b})\geq0$ for $M_3\leq0$ since $M_2=M_{\rm wh}<0$ is assumed.
Since Eq.~(\ref{Dis}) is satisfied, the domain of $V_2\leq0$ is 
$R_2\geq R_+$ and $0\leq R_2 \leq R_-$, and furthermore, we find $R_-\leq R_{\rm b}\leq R_+$ by the inequality 
$V_2(R_{\rm b})\geq0$. The constraint (\ref{r-cons-out}) implies that the only domain of $R_2\geq R_+$ is 
allowed for the motion of the shell-2. In the latter case, since $M_3$ is necessarily larger than $M_2$, 
we have
\begin{equation}
V_2(R_{\rm b})=1-2M_3\left(\frac{m_2^2}{2|M_3-M_2|}\right)^{-1}=1-\frac{2M_3}{R_{\rm b}}.
\end{equation}
Hence, if $R_{\rm b}$ is larger than or equal to $2M_3$, $V_2(R_{\rm b})>0$ and hence the situation is 
similar to the former case: The allowed domain for the shell-2 is $R_2\geq R_+$. 
As mentioned in the footnote 1, if $R_{\rm b}$ is smaller than $2M_3$, 
the allowed domain for the shell-2 is determined by the only condition $V_2 \leq0$. 

To summarize, one of the following three conditions should be satisfied so 
that the shell-2 falling from infinity enters the wormhole. 
By using the relation $M_2=M_{\rm wh}$ and 
$M_3=M_{\rm wh}+m_2\sqrt{E}$, 
\begin{itemize}
\item[E1)] $E\geq1$, if $M_{\rm wh}+m_2\sqrt{E}/2 \geq0$.
\item[E2)] $E>1$ and $1+4M_{\rm wh}(M_{\rm wh}+m_2\sqrt{E})/m_2^2<0$, if $M_{\rm wh}+m_2\sqrt{E}/2 <0$.
\item[E3)] $E>1$, $1+4M_{\rm wh}(M_{\rm wh}+m_2\sqrt{E})/m_2^2\geq0$ and $a>R_+$, 
if $M_{\rm wh}+m_2\sqrt{E}/2<0$. 
\end{itemize}

\section{Collision between the shells}

\begin{figure}
\begin{center}
\includegraphics[width=0.3\textwidth]{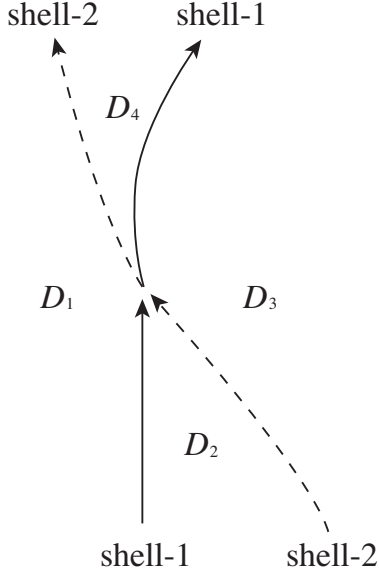}
\caption{\label{collision}
The shell-1 supporting the wormhole is initially static. The shell-2 falls into the wormhole and 
collides with the shell-1. The interaction between these shells is assumed to be gravity only:  
The shells merely go through each other. 
}
\end{center}
\end{figure}

Let us consider a process in which the shell-2 shrinks and collides the shell-1 
which supports the wormhole. The situation may be recognized by Fig.~\ref{collision}. 
The collision occurs at $r=a$. 
Then, in this section, we show how the 
mass parameter in the domain between the shells changes by the collision. 

We assume that the interaction between these shells is gravity only. Thus, after the collision, these shells 
merely go through each other: 
The 4-velocities $u_i^\alpha$ ($i=1,2$) of the shells are continuous at the collision event, 
respectively. We assume that the proper mass $m_A$ of each 
shell does not change. 

In the domain $D_2$, we have two tetrad basis 
$(u_i^\alpha, n_i^\alpha, e_{(\theta)}^\alpha, e_{(\phi)}^\alpha)$, where $i=1,2$.  
We can express the 4-velocity $u_1^\alpha$ of the shell-1 by using the tetrad basis 
$(u_2^\alpha, n_2^\alpha, e_{(\theta)}^\alpha, e_{(\phi)}^\alpha)$, and 
converse is also possible;
\begin{eqnarray}
u_1^\alpha&=&\left[-u_2^\alpha u_{2\beta}+n_2^\alpha n_{2\beta}+e_{(\theta)}^\alpha e_{(\theta)\beta}
+e_{(\phi)}^\alpha e_{(\phi)\beta}\right]u_1^\beta
=-(u_1^\beta u_{2\beta})u_2^\alpha+(u_1^\beta n_{2\beta})n_2^\alpha , \\
u_2^\alpha&=&\left[-u_1^\alpha u_{1\beta}+n_1^\alpha n_{1\beta}+e_{(\theta)}^\alpha e_{(\theta)\beta}
+e_{(\phi)}^\alpha e_{(\phi)\beta}\right]u_2^\beta
=-(u_2^\beta u_{1\beta})u_1^\alpha+(u_2^\beta n_{1\beta})n_1^\alpha.
\end{eqnarray}
The components of $u_i^\alpha$ and $n_i^\alpha$ with respect to the coordinate basis 
in $D_2$, i.e., $(t_2,r,\theta,\phi)$, are given by
\begin{eqnarray}
u_1^\alpha&=&\left(\frac{1}{\sqrt{f_2}},0,0,0\right), \\
n_1^\alpha&=&\left(0,\sqrt{f_2},0,0\right), \\
u_2^\alpha&=&\left(\frac{1}{f_2}\sqrt{\dot{R}_2^2+f_2}, \dot{R}_2,0,0\right), \\
n_2^\alpha&=&\left(\frac{\dot{R}_2}{f_2},\sqrt{\dot{R}_2^2+f_2}, 0,0\right), 
\end{eqnarray}
where $f_2=f_2(a)$. Hence, we have
\begin{eqnarray}
u_1^\beta u_{2\beta}&=&u_2^\beta u_{1\beta}=-\sqrt{\frac{\dot{R}_2^2}{f_2}+1}, \\
u_1^\beta n_{2\beta}&=&-\frac{\dot{R}_2}{\sqrt{f_2}}, \\
u_2^\beta n_{1\beta}&=&\frac{\dot{R}_2}{\sqrt{f_2}}.
\end{eqnarray}

\subsection{Shell-1 after the collision}

The tetrad basis $(u_2^\alpha, n_2^\alpha, e_{(\theta)}^\alpha, e_{(\phi)}^\alpha)$ is 
available also in the domain $D_3$. The components of $u_2^\alpha$ and $n_2^\alpha$ 
with respect to the coordinate basis in $D_3$ are given by
\begin{eqnarray}
u_2^\alpha&=&\left(\frac{1}{f_3}\sqrt{\dot{R}_2^2+f_3}, \dot{R}_2,0,0\right), \\
n_2^\alpha&=&\left(\frac{\dot{R}_2}{f_3},\sqrt{\dot{R}_2^2+f_3}, 0,0\right), 
\end{eqnarray}
where $f_3=f_3(a)$. By using the above equations, we obtain 
the components of $u_1^\alpha$ with respect to the coordinate basis in $D_3$ as
\begin{eqnarray}
u_1^{t_3}&=&-(u_1^\beta u_{2\beta})u_2^{t_3}+(u_1^\beta n_{2\beta})n_2^{t_3}
=-(u_1^\beta u_{2\beta})\frac{1}{f_3}\sqrt{\dot{R}_2^2+f_3}
+(u_1^\beta n_{2\beta})\frac{\dot{R}_2}{f_3} \cr
&=&\frac{1}{f_3\sqrt{f_2}}\left[
\sqrt{(\dot{R}_2^2+f_2)(\dot{R}_2^2+f_3)}-\dot{R}_2^2
\right], \\
u_1^r&=&-(u_1^\beta u_{2\beta})u_2^{r}+(u_1^\beta n_{2\beta})n_2^{r}
=-(u_1^\beta u_{2\beta})\dot{R}_2+(u_1^\beta n_{2\beta})\sqrt{\dot{R}_2^2+f_3}\cr
&=&\frac{\dot{R}_2}{\sqrt{f_2}}\left(
\sqrt{\dot{R}_2^2+f_2}-\sqrt{\dot{R}_2^2+f_3}
\right), \label{ur1}\\
u_1^\theta&=&u_1^\phi=0.
\end{eqnarray}
The above components are regarded as those of the 4-velocity of the shell-1 
in the domain $D_3$ just after the collision event.  
By using Eqs.~(\ref{j-con-out-1}) and (\ref{ur1}), we have 
\begin{equation}
u^r_1=\frac{m_2\dot{R}_2}{a\sqrt{f_2}}.
\end{equation}
By taking the square of Eq.~(\ref{j-con-out-1}) and using Eq.~(\ref{e-eq-2}), we have
\begin{eqnarray}
\sqrt{(\dot{R}_2^2+f_2)(\dot{R}_2^2+f_3)}=\dot{R}_2^2+\frac{f_2+f_3}{2}
-\frac{1}{2}\left(\frac{m_2}{a}\right)^2 =E-\left(\frac{m_2}{2a}\right)^2.
\end{eqnarray}
The above equation implies
\begin{equation}
E>\left(\frac{m_2}{2a}\right)^2.
\end{equation}
Then, we have
\begin{equation}
u_1^{t_3}=\frac{1}{f_3\sqrt{f_2}}\left[
1-\frac{2M_{\rm d}}{a}-\frac{1}{2}\left(\frac{m_2}{a}\right)^2
\right].
\end{equation}
We can check that the normalization condition $-f_3 (u_1^{t_3})^2 +f_3^{-1}(u_1^r)^2=-1$ is satisfied. 

The above result implies that after the collision, the derivative of the areal radius 
of the shell-1 with respect to its proper time becomes
\begin{equation}
\dot{R}_1|_{\rm after}=\frac{m_2\dot{R}_2}{a\sqrt{f_2(a)}}.
\label{R_1-after}
\end{equation}
Since the shell-2 falls into the wormhole, $\dot{R}_2$ is negative. 
This fact implies that the shell-1 or equivalently the radius of the wormhole throat 
begin shrinking just after the shell-1 collides with the shell-2 if $m_2$ is positive.  By contrast, 
if $m_2$ is negative, the shell-1 start to expand after the shell-2 goes through the wormhole. 
This result implies that $m_2$ plays a role of not only the proper mass of the shell-2 but also 
the active gravitational mass of it.  

The domain between the shell-1 and the shell-2 after the collision is called $D_4$. 
By the symmetry, $D_4$ is also described by the Schwarzschild geometry 
with the mass parameter $M_4$. 
From the junction condition between $D_4$ and $D_3$, the shell-1
obeys the following equation just after the collision;
\begin{equation}
\dot{R}_1^2|_{\rm after}
=-1+\left(\frac{M_3-M_4}{m_1}\right)^2+\frac{M_3+M_4}{R_1}+\left(\frac{m_1}{2R_1}\right)^2.
\label{E-1-after}
\end{equation}
From the above equation and Eq.~(\ref{R_1-after}), we obtain
\begin{equation}
\frac{1}{f_2}\left(\frac{m_2}{a}\right)^2 V_2(a)=1-\left(\frac{M_3-M_4}{m_1}\right)^2
-\frac{M_3+M_4}{a}-\left(\frac{m_1}{2a}\right)^2. 
\label{m-eq-1}
\end{equation}

\subsection{Shell-2 after the collision}

Since the tetrad basis $(u_1^\alpha, n_1^\alpha, e_{(\theta)}^\alpha, e_{(\phi)}^\alpha)$ is 
available also in the domain $D_1$. 
By using Eqs.~(\ref{th-th-comp}), (\ref{t-dot}) and (\ref{K-1}), the components of $u_1^\alpha$ and $n_1^\alpha$ 
with respect to the coordinate basis in $D_1$ are given by
\begin{eqnarray}
u_1^\alpha&=&\left(-\frac{1}{\sqrt{f_1}},0,0,0\right), \\
n_1^\alpha&=&\left(0, -\sqrt{f_1}, 0,0\right), 
\end{eqnarray}
where $f_1=f_1(a)$. As already noted just below Eq.~(\ref{K-1}), 
the time component of $u_1^\alpha$ 
with respect to the coordinate basis in $D_1$ is negative. 

By using the above equations, we obtain 
the components of $u_2^\alpha$ with respect to the coordinate basis in $D_1$ as
\begin{eqnarray}
u_2^{t_1}&=&-(u_2^\beta u_{1\beta})u_1^{t_1}+(u_2^\beta n_{1\beta})n_1^{t_1}
=(u_2^\beta u_{1\beta})\frac{1}{\sqrt{f_1}}=-\frac{1}{f_1}\sqrt{\dot{R}_2^2+f_1}, \\
u_2^r&=&-(u_2^\beta u_{1\beta})u_1^{r}+(u_2^\beta n_{1\beta})n_1^{r}
=-(u_2^\beta n_{1\beta})\sqrt{f_1}=-\sqrt{\frac{f_1}{f_2}}\dot{R}_2=-\dot{R}_2,
\label{u_2^r}\\
u_2^\theta&=&u_2^\phi=0.
\end{eqnarray}
where we have used the symmetric condition $f_1(a)=f_2(a)$.  Since $\dot{R}_2$ is negative, 
the shell-2 begins expanding after the collision. This is a reasonable result because of the wormhole 
structure. 

From the junction condition between $D_1$ and $D_4$, we have
\begin{equation}
\dot{R}_2^2|_{\rm after}=-1+\left(\frac{M_1-M_4}{m_2}\right)^2+\frac{M_1+M_4}{R_2}
+\left(\frac{m_2}{2R_2}\right)^2.
\end{equation}
From Eq.~(\ref{u_2^r}), since $\dot{R}_2^2$ is unchanged by the collision, we have 
\begin{equation}
V_2(a)=1-\left(\frac{M_1-M_4}{m_2}\right)^2-\frac{M_1+M_4}{a}
-\left(\frac{m_2}{2a}\right)^2. \label{m-eq-2}
\end{equation}

\subsection{The mass parameter $M_4$ in $D_4$}

Equations (\ref{m-eq-1}) and (\ref{m-eq-2}) impose the following constrains 
on one unknown parameter $M_4$; 
\begin{eqnarray}
V_2(a)&=&\left(\frac{a}{m_2}\right)^2 \left(1-\frac{2M_2}{a}\right) 
\left[1-\left(\frac{M_3-M_4}{m_1}\right)^2
-\frac{M_3+M_4}{a}-\left(\frac{m_1}{2a}\right)^2\right], \label{m-eq-3}\\
V_2(a)&=&1-\left(\frac{M_1-M_4}{m_2}\right)^2-\frac{M_1+M_4}{a}
-\left(\frac{m_2}{2a}\right)^2, \label{m-eq-4}
\end{eqnarray}
and by the definition of $V_2$, i.e.,  Eq.~(\ref{V2-def}), we have
\begin{equation}
V_2(a)=1-\left(\frac{M_3-M_2}{m_2}\right)^2-\frac{M_2+M_3}{a}
-\left(\frac{m_2}{2a}\right)^2. \label{V2a}
\end{equation}

Since $M_1=M_2=M_{\rm wh}$, Eqs.~(\ref{m-eq-4}) and (\ref{V2a}) lead to
\begin{equation}
\left(\frac{M_1-M_3}{m_2}\right)^2+\frac{M_1+M_3}{a}=\left(\frac{M_1-M_4}{m_2}\right)^2+\frac{M_1+M_4}{a}.
\end{equation}
By solving the above equation with respect to $M_4$, we obtain two roots, 
$M_4=M_3$ and $M_4=2M_1-M_3-m_2^2/a$.

Since the shell-1 is static before the collision, $V_1(a)$ should vanish, and this condition leads to
\begin{equation}
\left(\frac{m_1}{2a}\right)^2=f_2(a).
\end{equation}
By using the above condition and Eq.~(\ref{m-eq-3}), we find that 
\begin{equation}
M_4=M_{\rm wh}-m_2\left(\sqrt{E}+\frac{m_2}{a}\right)
\end{equation}
is a solution, where we have used $M_1=M_2=M_{\rm wh}$ and $M_3=M_{\rm wh}+m_2\sqrt{E}$. 
Hence, after the collision, the wormhole becomes asymmetric. 
Asymmetric wormhole is necessarily metastable, and hence the wormhole 
might collapse.   

\section{The condition that the wormhole persists}

In this section, we consider the condition that the wormhole stably 
exists after the entrance of the shell-2. 
First of all, $a>2M_3$ should hold. If it is not the case, the wormhole is enclosed by an event horizon 
after the shell-2 enters a domain of $R_2\leq2M_3$.

\begin{figure}
\begin{center}
\includegraphics[width=0.8\textwidth]{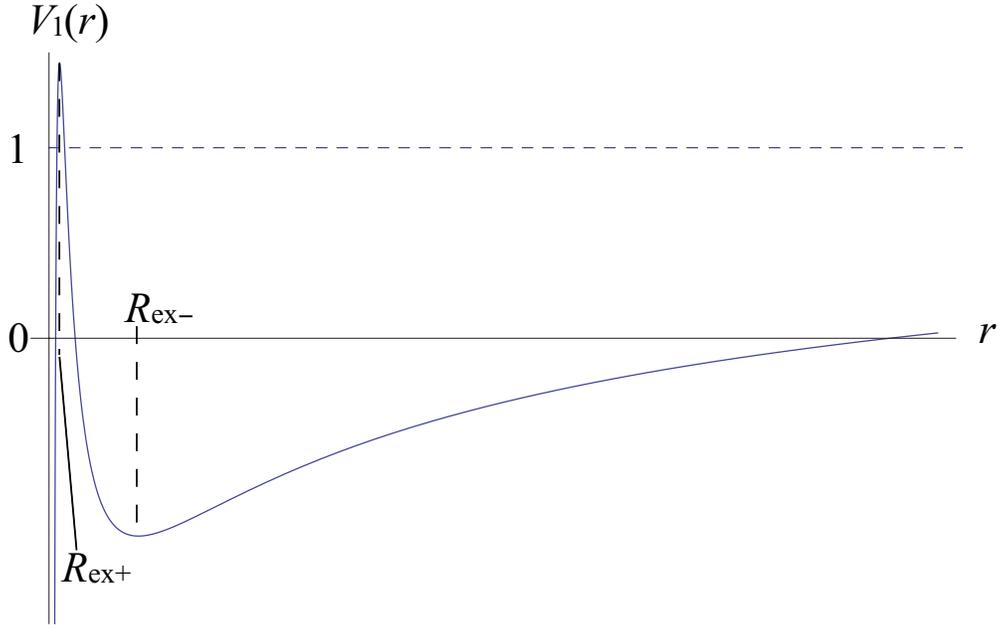}
\caption{\label{Vaf}
The effective potential of the shell-1 after the shell-2 goes through the wormhole. 
In this case, the wormhole persists. 
}
\end{center}
\end{figure}

Second, the effective potential of the shell-1 should have a negative minimum 
between positive potential domains. From Eq.~(\ref{E-1-after}), 
we see that the effective potential  of the shell-1 after the collision $\bar{V}_1$ 
is given by
\begin{eqnarray}
\bar{V}_1(r)
=1-\bar{\cal E}r^{4w}-\frac{2\bar{M}_{\rm wh}}{r}-\left(\frac{\mu}{2}\right)^2r^{-2(2w+1)},
\end{eqnarray}
where
\begin{eqnarray}
\bar{\cal E}&=&\left(\frac{M_4-M_3}{\mu}\right)^2=\left(\frac{m_2}{\mu}\right)^2
\left(2\sqrt{E}+\frac{m_2}{a}\right)^2, \\
&&\cr
\bar{M}_{\rm wh}&=&\frac{M_3+M_4}{2}=M_{\rm wh}-\frac{m_2^2}{2a}.
\end{eqnarray}
 By the condition (\ref{wh-con}), $\bar{V}_1(r)\rightarrow -\infty$ for $r\rightarrow0$, whereas 
 $\bar{V}_1(r)\rightarrow 1$ for $r\rightarrow\infty$. 
 Hence, the effective potential $\bar{V}_1$ should have at least 
 two extremums (see Fig.~\ref{Vaf}). 
The equation $\bar{V}'_1(r)=0$ is rewritten in the form
\begin{equation}
-4w\bar{\cal E}x^2+2\bar{M}_{\rm wh}x+\frac{1}{2}(2w+1)\mu^2=0,
\end{equation}
where $x=r^{4w+1}$.
Since, as mentioned, the effective potential $\bar{V}_1$ should 
have two extremum, the discriminant of the 
above quadratic equation should be positive, i.e., 
\begin{equation}
\bar{M}_{\rm wh}^2+2w(2w+1)\bar{\cal E}\mu^2>0.
\end{equation}
The two real roots of $\bar{V}_1'(r)=0$ is given by
\begin{equation}
r=R_{\rm ex\pm}=\left[-\frac{1}{4w\bar{\cal E}}\left(-\bar{M}_{\rm wh}
\pm\sqrt{\bar{M}_{\rm wh}^2+2w(2w+1)\bar{\cal E}\mu^2}\right)\right]^{1\over 4w+1}.
\end{equation}
The maximum of the effective potential $\bar{V}_1$ 
is at $r=R_{\rm ex+}$, and the condition $\bar{V}_1(R_{\rm ex+})>0$ 
should hold. Furthermore, the radius of the wormhole at 
the moment of the collision should be larger than $R_{\rm ex+}$. If so, 
$\bar{V}_1(R_{\rm ex-})$ is necessarily negative, and hence we need not impose 
$\bar{V}_1(R_{\rm ex-})<0$ in addition to $a>R_{\rm ex+}$. 

To summarize, all of the following four conditions should be satisfied 
so that the wormhole persists after the shell-2 goes through it. 
\begin{itemize}
\item[P1)] $a>2M_3$.
\item[P2)] $\bar{M}_{\rm wh}^2+2w(2w+1)\bar{\cal E}\mu^2>0$.
\item[P3)] $a>R_{\rm ex+}$.
\item[P4)] 
$\bar{V}_1(R_{\rm ex+})>0$
\end{itemize}

There are three independent parameters in this model. 
The initial static wormhole 
is characterized by the  constant of proportionality in the equation of state, $w$, 
and its gravitational mass, $M_{\rm wh}$: Note that $\mu$ and $a$ are the 
functions of $w$ and $M_{\rm wh}$ by Eqs.~(\ref{mu-sol}) and (\ref{a-sol}). 
However since $M_{\rm wh}$ may be regarded as a unit, 
the remaining parameter is only $w$, whereas the dust shell is 
characterized by two parameters, its proper mass $m_2$ and the square of 
conserved specific energy $E$.

\begin{figure}
\begin{center}
\includegraphics[width=0.5\textwidth]{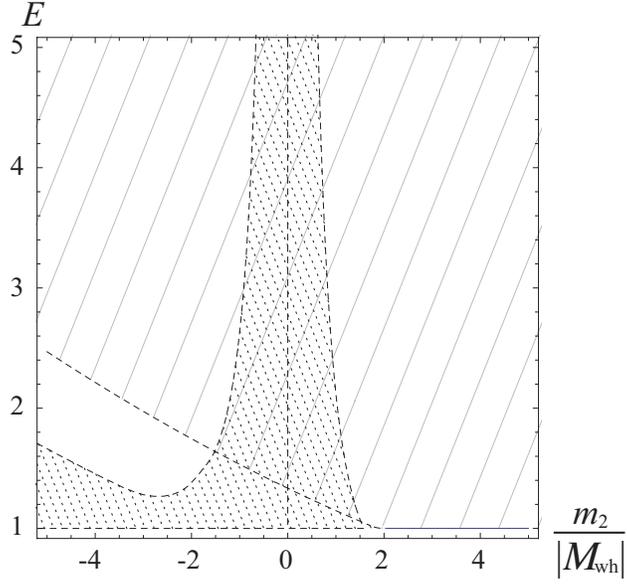}
\caption{\label{7-16}
The vertical axis represents $E$, whereas the horizontal axis is 
$m_2/|M_{\rm wh}|$. We assume $w=-7/16$.  
At least one of E1), E2) and E3) is satisfied in the domain shaded by straight lines. 
The domain shaded by dots is the intersection of the domains each of which 
P1)--P4) are satisfied. 
If the parameters $E$ and $m_2$ take values in the 
intersection of the domains shaded by straight lines and dots,
the wormhole does not 
collapse but merely oscillates after the shell-2 goes through the wormhole. 
}
\end{center}
\end{figure}

\begin{figure}
\begin{center}
\includegraphics[width=0.5\textwidth]{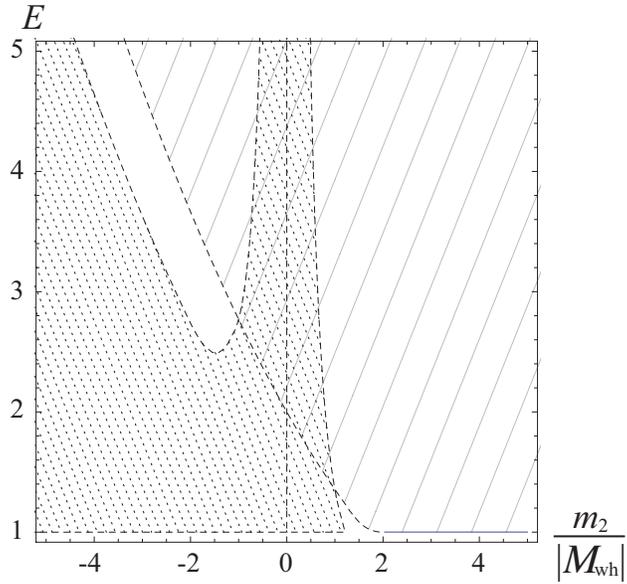}
\caption{\label{3-8}
The same as Fig.~\ref{7-16} but $w=-3/8$. 
}
\end{center}
\end{figure}

\begin{figure}
\begin{center}
\includegraphics[width=0.5\textwidth]{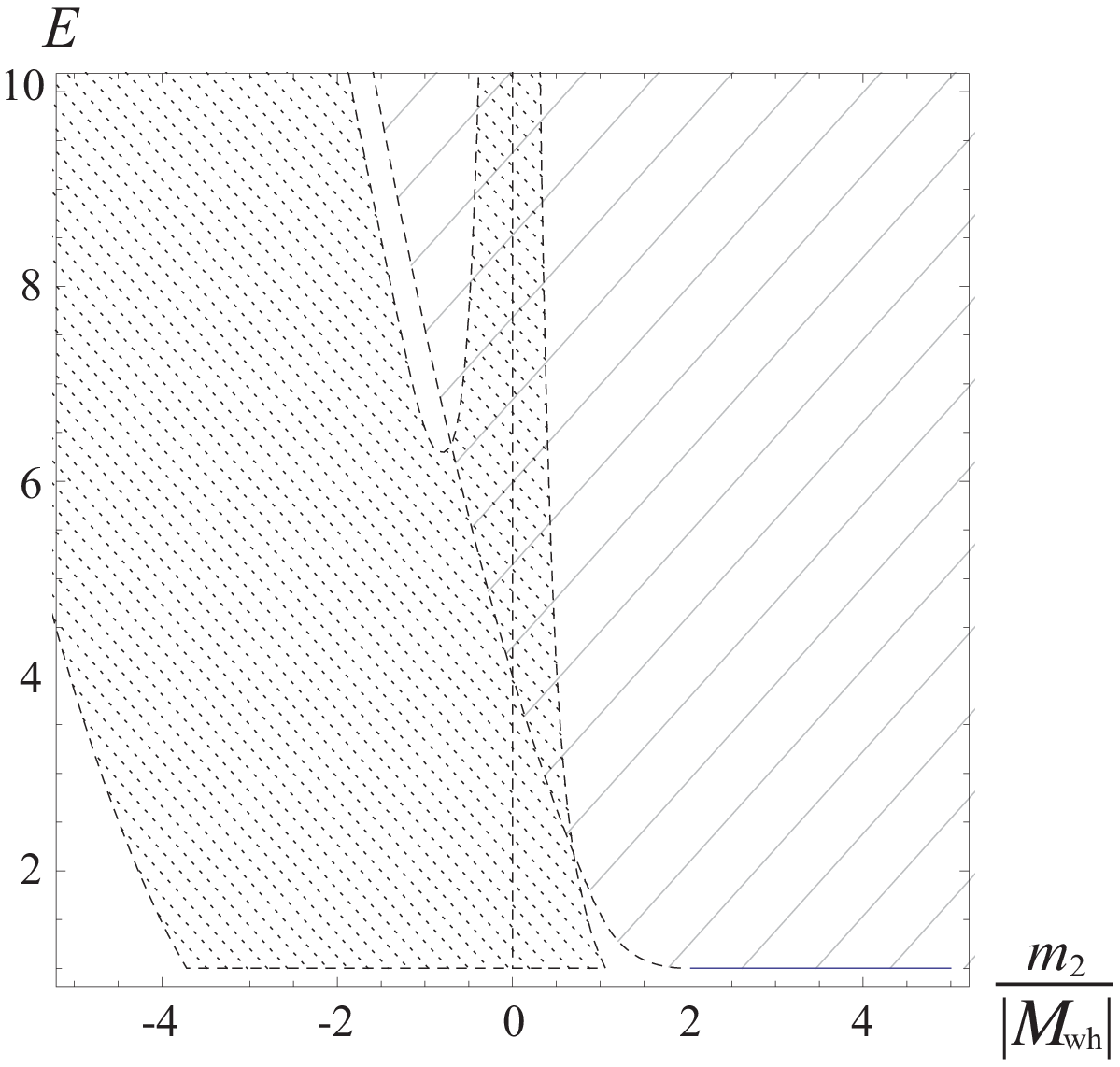}
\caption{\label{5-16}
The same as Fig.~\ref{7-16} but $w=-5/16$. 
}
\end{center}
\end{figure}

In Figs.~\ref{7-16}--\ref{5-16}, we depict the domains that satisfy 
the conditions of the entrance of the shell-2 
to the wormhole, E1)--E3), and 
the conditions of the persistence of the 
wormhole, P1)--P4), in $(E,m_2)$-plane in three cases of $w=-7/16,~ -3/8,~-5/16$, 
respectively.  
In the domain shaded by straight lines, at least one of E1), E2) and E3) is satisfied. 
The domain shaded by dots is the intersection of the domains each of which 
P1)--P4) are satisfied. If the parameters $E$ and $m_2$ take values in the 
intersection of the domains shaded by straight lines and dots, the wormhole does not 
collapse but merely oscillates after the shell-2 goes through the wormhole.

\section{Summary and discussion}

We analytically studied the dynamical process in which a spherical thin shell of dust goes through 
a wormhole supported by a spherical thin shell composed of the matter whose tangential pressure 
is proportional to its surface energy density with a constant of proportionality $w$.  
We treated these thin shells by Israel's formalism of metric junction. 

The negativ surface energy density of the shell is necessary to form the wormhole structure. 
This result is consistent with the known fact that the wormhole structure needs the violation of the 
null energy condition. 
We considered the situation in which the wormhole is 
initially static and has $Z^2$ symmetry with respect to the spherical thin shell supporting it, 
and found that the gravitational mass of the static wormhole should be negative, and 
the constant of proportionality in the equation of state 
should satisfies $-1/2<w<-1/4$, in order that the wormhole is stable 
against linear perturbations. 

Then we studied the condition that the wormhole persists  
after a spherical thin dust shell concentric with it goes through it.  
We assumed that the interaction between the wormhole shell and the dust shell is only gravity, 
or in other words, the 4-velocities of these shells are assumed to be continuous at the collision event. 
In this model, there are three free parameters: 
The  constant of proportionality, $w$, which characterizes the wormhole shell,  
the square of conserved specific energy $E$ and the proper mass $m_2$, 
which characterize the dust shell, in the unit 
that the initial gravitational mass of the wormhole is one. Then, we showed that  
there is a domain of the non-zero measure in $(m_2, E)$-plane for three values of $w$, 
in which the wormhole persists after the dust shell goes through it. 

In this paper, we investigated the case of the only linear equation of state for the shell 
supporting the wormhole. We need to investigate whether the present result strongly depends on 
the equation of state. This will be discussed elsewhere. 

\section*{Acknowledgments}
KN thanks the participants of ``workshop on theories and possibilities of observations 
of wormholes" held at Rikkyo university  in October 2012 for useful discussions.

\end{document}